\documentstyle[epsf,twocolumn,aps]{revtex}

\title{Optimal Mutation Rates in Dynamic Environments}

\author{Martin Nilsson}
\address
{Santa Fe Institute, 1399 Hyde Park Road, Santa Fe, New Mexico 87501 USA \\
Institute of Theoretical Physics, Chalmers University of Technology and G\"{o}teborg University, S-412 96 
G\"oteborg, Sweden {\tt martin@fy.chalmers.se}}

\author{Nigel Snoad}
\address{
Santa Fe Institute, 1399 Hyde Park Road, Santa Fe, New Mexico 87501 USA \\
The Australian National University, ACT 0200, Australia {\tt nigel@santafe.edu}}
\date{\today}

\begin{document}

\maketitle

\begin{abstract}
In this paper we study the evolution of the mutation rate for simple organisms in dynamic environments. A model with 
multiple fitness coding loci tracking a moving fitness peak is developed and an analytical expression for the 
optimal mutation rate is derived. Surprisingly it turns out that the optimal mutation rate per genome is approximately
independent of genome length, something that also has been observed in nature. Simulations confirm the 
theoretical predictions. We also suggest an explanation for the difference in mutation
frequency between RNA and DNA based organisms.
\end{abstract}

%\begin{multicols}{2}
\narrowtext

\section{Introduction}
%\multicolumn{2}{c}{test}
In any given environment the vast majority of mutations 
that have any effect on the fitness of a biological organism are deleterious.
One might expect the damaging effect of non-zero mutation rates 
to imply that when under evolutionary control 
the lowest  mutation rate compatible with physiological constraints should be selected for. 
However, when examined experimentally bacteria and viruses (and indeed all 
organisms) have 
significant non-zero rate, the magnitude and diversity of which
have failed to find satisfactory theoretical explanation.
Some results from a number of experiments measuring the mutation 
rates of a selection of small DNA-based organisms are shown
in Table~\ref{table:DNArates}.

\begin{table}
    \centering
\begin{tabular}{|lccc|} \hline\hline
Organism & $\nu$ & $\mu _b$ & $\mu _G$ \\\hline\hline 
Bacteriophage $M13$ & $6.4 \cdot 10^3 $ & $7.2 \cdot 10^{-7}$ & 
0.0046  \\ 
Bacteriophage $\l$ & $4.9 \cdot 10^4 $ & $7.7 \cdot 10^{-8}$ & $ 
0.0038 $ \\
Bacteriophage $T2$ \& $T4$ & $1.7 \cdot 10^5 $ & $2.4 \cdot 10^{-8}$ 
& $ 0.0040 $ \\
{\it E. coli} & $4.6 \cdot 10^6 $ & $4.1 \cdot 10^{-10}$ & $ 0.0025 
$ \\
{\it S. cerevisiae} & $1.2 \cdot 10^7 $ & $2.2 \cdot 10^{-10}$ & $ 
0.0027 $ \\
{\it N. crassa} & $4.2 \cdot 10^7 $ & $7.2 \cdot 10^{-11}$ & $ 0.0030 
$  \\ \hline\hline
\end{tabular}\label{table:DNArates}
\caption[table1]{Spontaneous mutation rates (per base $\mu_b$ and per 
genome 
$\mu _G$) in DNA-based microbes with different genome lengths $\nu$. 
(Data 
reproduced from Drake \textit{et al.}~\cite{DCCC98})} 
\nopagebreak
\end{table}
\nopagebreak

Despite the huge variation in 
genome length over four orders of magnitude the mutation rate per genome 
and replication $\mu_{G}$ remains
constant roughly within a factor of roughly $2$ (which is at the same level 
as the estimated accuracy of the figures). 
As pointed out by Drake and others \cite{DCCC98,M-SS95} this constancy in $\mu _{G}$ is 
surprising since DNA/RNA repair and transcription 
are primarily local processes that act on individual bases. Thus the data strongly suggest that 
point mutation rates for the different organisms have evolved towards individual optimal values that  
result in almost constant genomic copying fidelity.  

In this paper we develop a model of the evolution of mutation rates based on changing environments. 
The evolved point mutation rate of this model scales so that the genomic copying fidelity 
is approximately independent of genome length and insensitive to other parameters in the model. The evolved
mutation rates are also of the same
magnitude as observed in Table~\ref{table:DNArates} for biologically plausible parameter settings.
We also suggest a possible explanation for the high mutation rates of RNA viruses. Simulations confirm the 
predictions of the model.

%%%%%%%%%%%%%%%%%%%%%%

\section{Evolving Mutation Rates}

It is impossible to perfectly maintain and 
copy genetic information. All molecules, including DNA and RNA  are thermodynamically unstable, and 
their physical structure and hence the 
information they encode changes over time. In addition the binding sites of enzymes such 
as DNA polymerase are not perfectly specific and errors will be 
introduced during replication. Lowering 
the error rate requires the use of increasingly complex proof-reading 
and repair mechanisms, all of which ultimately impose an energetic, and 
hence fitness, cost on the organism. We can expect 
a balance to develop between the pressure to lower mutation rates due 
to the fitness cost of deleterious mutants and the physiological cost of 
high copying accuracy \cite{Kimura67,Leigh70,Kondrashov95}. Such a balance certainly
provides an ultimate lower limit to the mutation rate of all 
organisms but explaining the concstancy in genomic copying fidelity using such
arguments causes unnatural assumptions on the relation between cost of local copying fidelity
and genome length. There is also little experimental evidence that mutation rates are actually 
determined by such a balance.

When viewed as a whole the genome encodes not only proteins that 
directly influence its reproductive or survival ability,
but also  the copying fidelity  with which the genome 
reproduces. For example some viroids contain genes that are translated into surface coat 
proteins while others genes code for the replicase enzymes that perform the copying of its genetic 
material. In more complex organisms
additional genes may encode for modifiers of the accuracy of copy and repair enzymes, usually
 increasing mutation rates 
\cite{Cox74,Cox76a,Chao83,SGL97}, but sometimes resulting in  a decrease 
\cite{MH97}. These modifiers 
can have large or small effects on mutation rate and affect individual bases or the entire genome
\cite{Moxon..94,MT97,Radman99}.

One consequence of this flexibility of mutation rates and their encoding is that if there are random 
changes (mutations) in genes determining the mutation rate then the copying fidelity
will itself undergo Darwinian evolution.

%%%%%%%%%%%%%%%%%%%%%%%%%%%%%%%%%%%%%%%%%%%%%%%%%%%%%%%%%
\section{Population Genetics in Changing Environments}
\label{sec:optreview}

When comparing two hap\-loid
gen\-omes, the one with lower mutation 
frequency will produce offspring that are on 
average more closely related to itself. This means that for an asexual haploid replicator evolving on a static 
fitness landscape the optimal mutation rate for a sequence whose 
fitness is already globally maximal is zero. If the fitness peak moves, however, 
the situation changes: to avoid extinction a
 genome with an initially superior fitness is forced to accept a non-zero mutation 
rate to survive. This leads to a non-trivial optimal copying fidelity.

Kimura formalized the evolutionary effect of a changing 
environment by considering the genetic load of a population~\cite{Kimura67}: the proportion by 
which the population fitness is decreased in comparison with an 
optimum genotype. Genetic load results from a number of competing 
factors; most notably the \emph{mutational load} due to the 
deleterious effects of most mutations  and the
\emph{segregational load} due to the temporary reduction in fitness that occurs 
whenever the selective environment changes. Assuming 
that a population minimizes the genetic load, the optimal mutation rate can be calculated.
Using a descrete time model, i.e. a model where there is no
overlap between generations, with one fitness determining locus the optimal mutation rate becomes:

\begin{equation}\label{eqn:leighopt}
    \mu_{opt}=\frac{1}{\tau}
\end{equation}
where $\tau$ is the number of generations between environmental changes. 
This model only considers 
the effect of mutations on the population and is therefore based on group selection.  

Later population genetic models that examined competition between 
genetic modifiers of the mutation rate demonstrated that (for 
haploids with a single fitness determining locus) a non-zero mutation rate comes to dominate a population 
evolving in an oscillating environment~\cite{Leigh73,Gillespie81,Ishii..89,Gillespie91}.  
These models are not built on group selection. However a general and simple to interpret multi-locus 
modifier model does not exist.

%%%%%%%%%%%%%%%%%%%%%%%%%%%%%%%%%%%%%%%%%%%%%%%%%%%%%%%%%

\section{The Model}
\label{sec:optgeneral}

We will explore a more general model of the evolution of mutation rates in a dynamic environment.
Consider a population of haploid genomes where a genome consists of two separated parts, one 
coding for the fitness and one coding for the probability per base $\mu$ of an error occurring during 
copying. There is complete linkage (no recombination)
between the sections of the genome that encode the mutation rate and those that determine the fitness. 
We also assume that the fitness determining region is of fixed length $\nu$. In general we are interested
in the fates of certain genomes $g_i$ which have a (possibly time-dependent) fitness advantage 
$\sigma(t)$ over all other sequences. We call these genomes master-sequences. The genomic copying fidelity 
of the fitness determining region  of each strain $g_i$ is $Q_i=(1-\mu _i)^\nu$, the index $i$ 
refers to the mutation rate of the strain, different strains have different mutation rates but identical
fitness $\sigma$.
We assume that mutations do not affect the copying fidelity, only the fitness. Changes to the mutation 
rates occur on a time-scale significantly slower 
than the time it takes for the population to reach equilibrium.  During a period when a specific 
sequence has superior fitness compared to the background (i.e. between environmental shifts)
the  changes in the relative concentrations $x_i$ of the master-sequences are described by the 
replicator equation
\begin{eqnarray}
	\dot{x} _i (t) & = & Q_i \sigma (t) x_i(t) - f(t) x_i (t) 	
\label{eqn:rate-eq}
\end{eqnarray}
where $f(t) = \sigma (t) \sum _j Q_j x_j (t)$ normalizes the relative
concentrations of the master-sequence strains. Mutations from background sequences onto the strains 
with optimal fitness are ignored. Since we are only interested in competition between  
master-sequences the background is not explicitly expressed in these equations. 

The environment changes as follows: for a time $t \in [ 0 , \tau _1 ]$ one genotype has superior fitness, 
followed by a new gene-sequence for time $t \in [ \tau _1 , \tau _1 + \tau _2 ]$, etc.  
The notation is chosen so that $\tau$ denotes lengths of time intervalls.
We  assume that the initial concentration of the new 
master-sequences $x_i$ immediately after the shift 
(at time $t _a = \sum _{i=1}^m \tau _i + \epsilon$, where $m$ denotes shifts of the fitness-peak
and $\epsilon$ is am infinitely small time-period)
 are proportional to the concentrations of the old master-sequence
before the shift (at $t _b = \sum _{i=1}^m \tau _i  - \epsilon$)
\begin{eqnarray}
	x_{i} (t_a) & = & h (\mu _i) x_{i} (t_b)
\end{eqnarray}
It is reasonable to assume that  $h (\mu _i)$ is a function with Taylor-expansion
in the mutation rate $\mu$
\begin{eqnarray}
	h (\mu) & = & \sum _{j=k_m} ^{\infty} \alpha _j \mu ^j 
\label{taylor}
\end{eqnarray}
where $k_m$ is a measure of the environmental change, i.e. the number of point mutations needed to transform the old 
superior sequence into the new. This basically means that $k_m$ is the Hamming distance from the old 
peak to the new at shift $m$. The constants $\alpha _j$ are combinatorial factors. It will turn out
that the optimal mutation rate is independent of these factors.

To analyze the long term behavior of this system we make 
a change of variables $y_{i} (t) = e^{\int ^t_0 f (s) d s} x_{i} (t)$. The
new system of differential equations is linear and the
equations are decoupled (due to the assumption that the 
selective dynamics is significantly faster than the changes in mutation 
rate), it is therefore easy to find the analytical solution:
\begin{eqnarray}
	y_{i} (t) & = & y_{i} (0) e^{Q_i \int _0 ^t \sigma _m (s) d s}
\end{eqnarray}

Since $x_i$ is propotional to $y_i$, maximizing the growth of $y_i$ and $x_i$ are equivalent.
After a suitably long time interval the population will be completely dominated by 
genomes that have a mutation
rate closest to the optimal value $\mu _{opt}$ which maximizes the long 
term growth of the strain
\begin{equation}
	\max _{\mu} \left( \Pi _m h (\mu) e ^{(1-\mu)^{\nu}  \langle \sigma 
\rangle _m \tau _m  } \right) 
\end{equation}
where $ \langle \cdot \rangle_m $ denotes a time average during time-period $m$.
Setting the derivative of this expression to zero and using Eq.~\ref{taylor} 
we find the optimal copying fidelity to be approximately

\begin{eqnarray}
\mu _{opt} & = & \frac{\langle k \rangle}{\nu \langle \sigma \rangle \langle \tau \rangle}
\end{eqnarray}
where $ \langle \cdot \rangle$ denotes a time average over all time periods. We also assume
no correlation between $\langle \sigma \rangle _m$ and $\tau _m$.
Since the genome lengths is large $\nu \gg 1$, the optimal copying fidelity and mutation rate 
per genome become:
\begin{eqnarray}
	Q_{opt} & = &   e^{- \frac{\langle k \rangle}{\langle \sigma 
	\rangle \langle \tau \rangle}} \label{eqn:Qopt}\\
 	\mu _{G} & = & \frac{\langle k \rangle}{\langle \sigma 
	\rangle \langle \tau \rangle} \label{eqn:muG}
\end{eqnarray}

Thus we find that the genomic optimal copying fidelity is independent of the genome 
length for fairly general types of environmental change in both the advantage of the fittest genotype 
$\sigma (t)$ and the size of environmental shifts $h (\mu)$. 

%%%%%%%%%%%%%%%%%%%%%%%%%%%%%%%%%%%%
\section{Simulations}
\label{sec:optsimulations}

To confirm the theoretical derivations we simulated
the evolution of replicators in continuous time on a moving single peaked landscape 
using a birth-death process.
Each time unit in the  continuous time replicator equation is
the mean replacement time of the population and could therefore 
be identified as a generation. In the simulation each 
generation is devided into $N$ time-steps (where $N$ is the population size). At each 
of these time-steps a single individual is selected to copy and mutate. 
Individuals are selected wita h probability proportional to their 
relative fitness, which is given by $\sigma$ or $1$ on the single-peaked 
landscape. Thus a master-sequence of strain $g_i$ (with mutation rate $\mu _i$) 
is chosen with probability $\frac{x_i \sigma}{\langle f \rangle}$. This copy replaces
a randomly  chosen individual 
in the existing population which is then discarded. Thus the population is 
replaced one by one in discrete birth-death events. In the limit of large 
population size the dynamics of this simulation 
approaches the continuous time replicator equation.

The fitness peak is changed every $\tau$ 
generations to one of its nearest neighbors. For the binary genomes used here 
it accomplished by flipping a randomly chosen bit in the definition of the fitness peak. 

\begin{figure}[l]
\centering
\leavevmode
\epsfxsize = 0.8 \columnwidth
\epsfbox{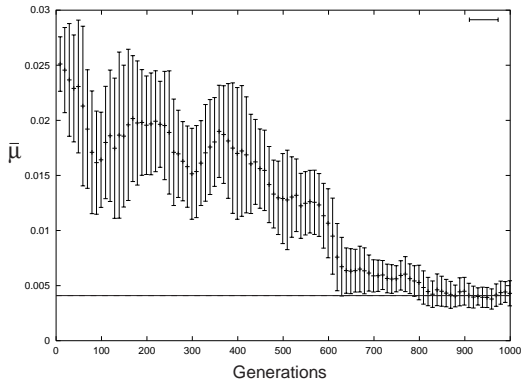}
\caption{\small Mean mutation rate evolving towards the optimal rate of 
$\mu _{opt}=0.00445$. Error bars are one standard deviation about the 
mean. $\sigma =5,\tau =2,\nu =25,N=10^4$ }
\label{fig:711mut}
\end{figure}

The population was first seeded with a diverse range of mutation rates and the 
population was allowed to evolve while these rates were kept fixed. This is a true test of $\mu _{opt}$,
since the fastest growing sequence should
come to dominate. In general the population 
converged to the strain with mutation rate closest to the theoretically predicted $\mu _{opt}$.  
Figure~\ref{fig:711mut} shows the mean mutation rate of the population $\bar{\mu}$ 
evolving down towards 
the theoretically predicted optimum $\mu _{opt} \approx \frac{1}{\nu \sigma \tau} = 0.00445$. 
From about generation 800 the variance in mutation rates in the population is 
larger than the fluctuations in the mean and the evolution of rates has effectively ended.

Simulations were also made to study the effects of more rapidly 
changing mutator dynamics. In these simulations errors in the copying process 
not only introduce changes in the fitness determining genotype, but 
also result in offspring with slightly different mutation rates than 
their parents, i.e. the mutation rate is allowed to evolve. The mutation rate was treated 
as a continuous variable which had Gaussian noise introduced during the copying process.

\begin{figure}[tp]
\centering
\leavevmode
\epsfxsize = 0.7 \columnwidth
\epsfbox{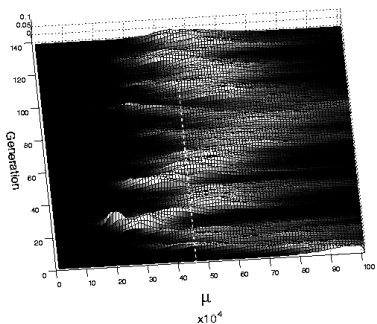}
\caption{\small Evolution of mutation rates of mutationally diverse population. 
$\mu _{opt}=4.45 \times 10^{-3}$, $\sigma=5,\tau =2,\nu =25,N=10^4$}
\label{fig:728dbn}
\end{figure}

Fig.~\ref{fig:728dbn} shows the evolution of mutation rates in detail in a population with 
a reasonably fast rate of change of mutation rates. This 
simulation has the same landscape parameters as Fig.~\ref{fig:711mut}. The mean mutation rate
fluctuates around the optimum. For mutation rates close 
to the optimum fluctuations in selection 
are significantly larger than the selective advantages of one mutation rate over 
another. In this region the 
evolution of mutation rates  is effectively neutral and thus the mean mutation 
rate conducts a random walk about the optimum. We also note that the population
 typically spends more time with mutation rates above the optimum than below.
This is mainly a finite population size effect. When the peak moves and the population size 
is limited there is a relatively large probability that there will be no individuals representing a
master-sequence with very low mutation rate on the new peak. This leads to a temporary increase
in mutation rate in the population after an environmental shift.

%%%%%%%%%%%%%%%%%%%%%%%%%%%%%%%%%%%%%%%%%%%%%%%%%%%%%%%
\section{Biological Implications}\label{sec:optimplications}

In nature the existence, and value, of an optimum mutation rate that results from a 
changing environment depends on many different  
parameters:  the time between shifts in the
selective environment, the
complex structure of the fitness-landscape, the genome length, co-evolutionary effects, the strength of 
selection, neutrality in the fitness landscape and fluctuations due to finite 
population sizes etc. One must therefore be careful when comparing the results of a simple model, 
such as the one we have presented in this paper, 
and biological measurements. Nonetheless it is this range of 
possible differences between organisms and the complexity of their evolutionary environments  that 
leads us to consider the possibility that simple laws 
of biology --- such as the scaling of point mutation rates with genome length  --- 
are likely to have quite simple explanations that do not depend 
on the details of the particular organism. It is therefore worth comparing the 
results of the model presented in this paper with the biological data.

\begin{figure}[p]
\centering
\leavevmode
\epsfxsize = 0.8 \columnwidth
\epsfbox{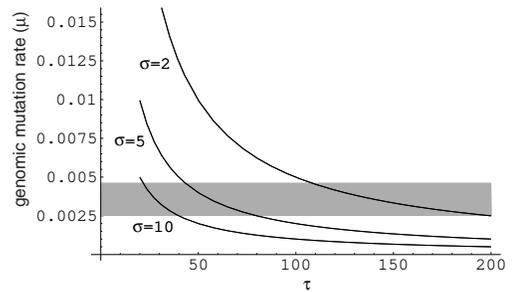}
\caption{\small The shaded region shows the genomic mutation rates for DNA based organisms 
listed in Table~\ref{table:DNArates}. For low average fitness advantage $\sigma$ the mutation rate
is relatively insensitive to the frequency of changes in the environment. For clarity we have assumed
$\langle k \rangle = 1$ in this figure.}
\label{fig:theory}
\end{figure}

For low mutation rates Eq.~\ref{eqn:muG} is relatively insensitive to changes in 
the average fitness or size and frequency of environmental changes, as shownin Fig.~\ref{fig:theory}.
This insensitivity of the optimal genomic mutation rates to evolutionary parameters  
is important, since the bacteria 
and phages illustrated in table~\ref{table:DNArates} are most unlikely to live 
in environments with the same types of time-dynamics and 
time-scales. In Fig.~\ref{fig:theory} we see that the sensitivity to one of the parameters in the 
model, $\sigma$ or $\tau$, depends strongly on in which region the other parameter is. For
most realistiv populations  we may expect the selective advantage $\sigma$ to be weak, maybe on average less
than $2$. The predicted mutation rate is then highly insensitive to the average time between shifts in the fitness 
landscape, e.g. $\sigma = 2$ gives $\tau \in [110 , 200 ]$ for the organisms listed in Table~\ref{table:DNArates}.
It is also reasonable to assume the fitness landscapes of the organisms listed in Table~\ref{table:DNArates}
to be more similar to each other than to higher eukaryotes and since our predictions as to $Q_{opt}$ are
 rather insensitive to the details of $\sigma (t)$, $\tau$  and $h (\mu)$ we would expect many organisms to have 
approximately the same mutation rate per genome (within an order of magnitude). This is what we observe 
for simple DNA-based organisms.

%%%%%%%%%%%%%%%%%%%%%%%%%%%%%%%%%%%%%%%%%%%%%%%%%%%
\section{RNA viruses}
\label{sec:coding}

The lytic RNA viruses consistently show an extremely high 
mutation rate --- orders of magnitude larger than that of any DNA viruses of similar size. This rate of 
around one substitution per genome per generation is inconsistent with the analysis conducted above
for mutation rates evolving in a changing selective environment. 
Such high rates imply implausible values for the dynamic environment parameters. 

As an explanation for the high mutation rates observed in many RNA viruses and the mutation rate scaling
with genome length it has been
suggested that these viruses have evolved the highest mutation rate possible to be able to adapt to a
rapidly changing environment. The maximal mutation rate is then given by the error-threshold, which was 
first discussed in a model by Eigen \textit{et al.}~\cite{Eigen77}. It basically states that on a 
singled peaked fitness landscape 
an organism must have high enough copying fidelity so that its relative superiority in reproduction rate multiplied by the 
probability of reproducing onto a perfect copy of itself must be larger than one, otherwise there will
be no effective selection for the genotype. It has later been shown that the error-threshold can rather easily be
generalized to include effects of a dynamic environment~\cite{NS99a}. From this argument it is however not clear
why RNA viruses should evolve towards the error-threshold while DNA based organism tend to have
much lower mutation rates (by orders of magnitude). In this section we will combine the 
error-threshold with the model presented in this paper to suggest a possible explanation to
the difference in observed mutation rates between DNA and RNA based organisms.

The dynamic environment model presented in this paper applies to organisms where the copying 
fidelity is encoded in a part of the genome that has little or no effect on fitness. In many viruses 
this may not be appropriate, partly because the  proteins involved in mutagenesis may have a multitude of
functions but also because the relatively high selective pressure towards short genome 
lengths will result in the overlap and multiple use of genetic 
material where possible. This give rise to a different possibility for the evolution of optimal mutation 
rates and might help explain the large differences between the observations for RNA and DNA based organisms.

We suggest that for organisms which have strong overlaps between genes coding for the mutation rate
and genes coding more directly for reproductive advantage there is no effective
selection for lower mutation rates, as long as the mutation rate is below the error threshold. This
argument is based on the assumption that most mutations 
are deleterious in terms of fitness, and that the relative fitness advantage on the local peak
results in stronger selection pressure than the pressure towards lower mutation rates. We also assume that
evolution of mutation rates usually affect regions of the genome where the organism need
mutations to be able to adapt ot changes in the environment. If these assumptions apply
we expect a population to have mutation rates close to the error-threshold.
Changes to mutation rate is transient, 
assuming that the organism is not pushed beyond the error-threshold. 

For this hypotheses to apply, viruses with high mutation rate 
(mainly RNA viruses) should have overlapping genes regulating mutation frequency as well as reproduction rate, 
whereas organisms with low mutation rates (such as those listed in Table~\ref{table:DNArates})
should not have overlapping reading frames in their genomes. There are observations that support
this, but it is unclear whether the correlation is strong enough for this hypothesis to be valid.

%%%%%%%%%%%%%%%%%%%%%%%%%%%%%%%
\section{Conclusions}
\label{sec:optconclusion}

In this paper we have studied the evolution of mutation rates in a population of multi locus genomes.
The genomic mutation rate $\mu _G$ leading to the greatest long term growth of a strain (the optimal rate) 
was analytically determined for reasonably general peak shifts and time-dependent replication rates $\sigma (t)$

\begin{displaymath}
    \mu _{G}\approx\frac{\langle k\rangle}{\langle \sigma \rangle \langle \tau \rangle}
\end{displaymath}
where $\langle k\rangle$ is the mean Hamming distance between successive fitness optima and 
$\langle \tau \rangle$ is the mean time between shifts.
 These optimal rates were quantitatively confirmed by computational simulations of 
populations whose mutation rates were allowed to evolve.
 
These continuous time multi-locus replicator models predict the kind of scaling of point-mutation rate 
with genome length that has been observed in some bacteria and viruses/phages and puzzled over for years. 
When combined with the consequences of the multiple use/pleiotropic encoding of copying machinery these models of 
the evolution of mutation rate in dynamic environments also suggest why lytic RNA viruses may have 
rates at or about the error-threshold. 

We would like to thank Claes Andersson and Erik van Nimwegen for useful discussions. 
Thanks are also due to Mats Nordahl 
who has given valuable comments on the manuscript. Nigel Snoad and Martin Nilsson were supported by 
SFI core funding grants. N.S. would also like to acknowledge the support of Marc Feldman and the 
Center for Computational Genetics 
and Biological Modeling at Stanford University while preparing this manuscript.

\bibliographystyle{unsrt}

%\end{multicols}
\end{document}